\documentclass[sigconf]{acmart}

\usepackage{subfig}
\usepackage{array}
\newcolumntype{L}{>{\centering\arraybackslash}m{0.4in}}

\AtBeginDocument{%
  \providecommand\BibTeX{{%
    \normalfont B\kern-0.5em{\scshape i\kern-0.25em b}\kern-0.8em\TeX}}}

\setcopyright{acmcopyright}
\copyrightyear{2022}
\acmYear{2022}
\setcopyright{rightsretained}
\acmConference[ICCAD '22]{IEEE/ACM International Conference on Computer-Aided Design}{October 30-November 3, 2022}{San Diego, CA, USA}
\acmBooktitle{IEEE/ACM International Conference on Computer-Aided Design (ICCAD '22), October 30-November 3, 2022, San Diego, CA, USA}
\acmDOI{10.1145/3508352.3549381}
\acmISBN{978-1-4503-9217-4/22/10}

\begin{document}

\title{Inhale: Enabling High-Performance and Energy-Efficient In-SRAM Cryptographic Hash for IoT}

\author{Jingyao Zhang}
\affiliation{%
  \institution{Department of Computer Science\\ University of California, Riverside}
  \country{USA}
  }
\email{jzhan502@ucr.edu}

\author{Elaheh Sadredini}
\affiliation{%
  \institution{Department of Computer Science\\ University of California, Riverside}
  \country{USA}
  }
\email{elaheh@cs.ucr.edu}

\renewcommand{\shortauthors}{Zhang et al.}

\begin{abstract}
In the age of big data, information security has become a major issue of debate, especially with the rise of the Internet of Things (IoT), where attackers can effortlessly obtain physical access to edge devices. The hash algorithm is the current foundation for data integrity and authentication. However, it is challenging to provide a high-performance, high-throughput, and energy-efficient solution on resource-constrained edge devices. In this paper, we propose \textit{Inhale}, an in-SRAM architecture to effectively compute hash algorithms with innovative data alignment and efficient read/write strategies to implicitly execute data shift operations through the in-situ controller.
We present two variations of \textit{Inhale}: \textit{Inhale-Opt}, which is optimized for latency, throughput, and area-overhead; and \textit{Inhale-Flex}, which offers flexibility in repurposing a part of last-level caches for hash computation.  
We thoroughly evaluate our proposed architectures on both SRAM and ReRAM memories and compare them with the state-of-the-art in-memory and ASIC accelerators. Our performance evaluation confirms that \textit{Inhale} can achieve 1.4$\times$ - 14.5$\times$ higher throughput-per-area and about two-orders-of-magnitude higher throughput-per-area-per-energy compared to the state-of-the-art solutions. 
\end{abstract}




\maketitle

\section{Introduction}
Information security has long been an issue of significant concern and investment willingness. After entering the era of big data, hospitals, large corporations, and governments have exerted considerable effort to assure data security and integrity \cite{cawthra_data_2020}. The current algorithms behind data integrity and authentication are the hashing algorithms. For instance, the widely used Transport Layer Security protocol provides data integrity for identity authentication by utilizing the hash algorithm \cite{dierks_transport_2008}. In addition, six of the seven candidate algorithms in the final round of the latest NIST post-quantum cryptography standardization process are based on hash algorithms \cite{alagic_status_2020}, five of which are specifically based on the \textit{Keccak} secure hash algorithm \cite{dworkin_sha-3_2015} (the NIST winner for SHA-3), which accounts for almost 50\% of the total algorithm's cycles \cite{alagic_status_2020,avanzi_crystals-kyber_2019}.
Keccak is also heavily used in cryptocurrencies, such as Ethereum \cite{wood_ethereum_2014}, Monero \cite{moser_empirical_2017}, and NEM \cite{zhang_stylized_2018}. Therefore, providing a high-throughput and energy-efficient solution for Keccak is critical for miners.

Moreover, an increasing number of edge devices are becoming capable of device-to-device, device-to-cloud, and device-to-user communication \cite{madakam_internet_2015}. To prevent attackers from tampering with sensitive data, these interactions must be constructed on top of identity authentication, which is the ability of individual devices to determine each other's identity through certain processes \cite{mahmoud_internet_2015}. Numerous approaches have been presented to mitigate IoT security concerns \cite{won_decentralized_2018,obiri_sovereign_2021,singla_blockchain-based_2018}. Since all of these mechanisms rely on hash algorithms, there is an imminent need for a high-throughput, low-latency, and energy-efficient solutions for hash algorithms on resource-constrained edge devices.

Several hardware acceleration solutions are proposed to improve throughput, latency, or hardware overhead. Tillich et al. \cite{tillich_high-speed_2009} use small LUTs to store intermediate variables to reduce the latency. Akin et al. \cite{akin_efficient_2010} exploit a pipelined design to implement different stages of hashing algorithms to improve the throughput. Moreover, Pessl et al. \cite{pessl_pushing_2013} propose to reduce on-chip storage overhead by interleaving input data. However, it is extremely challenging to optimize for high-throughput, low-latency, energy-efficiency, and area-efficiency all at the same time.

Recent studies based on in-memory \cite{bhattacharjee_sha-3_2017,nagarajan_shine_2019,zhang_recryptor_2018,reis_imcrypto_2022,sealer} computing are proposed to improve the performance and energy efficiency of secure-related algorithms, especially hash algorithms. 
Nagarajan et al. \cite{nagarajan_shine_2019} propose a ReRAM-based implementation of SHA-3, named \textit{SHINE}. To optimize for latency, the authors present a pipeline design by replicating the memory arrays in each stage of the pipeline, thus, trading hardware overhead and throughput-per-unit-area for latency. Moreover, their design does not generalize to other hashing algorithms. In addition, it expands the trusted computing base (TCB) to the off-chip memory, which exposes more space to attackers.
Zhang et al. \cite{zhang_recryptor_2018} propose \textit{Recryptor}, a reconfigurable cryptographic processor capable of accelerating finite field multiplication and reduction, AES, and Keccak using in-memory computing. However, due to the complex peripheral circuitry, Recryptor only runs at 28.8 MHz, which significantly limits the performance.

In this paper, we present \textit{Inhale}, a high-performance, low-overhead, and energy-efficient hashing accelerator for low-end IoT devices by repurposing 6T SRAM subarrays and turning them into active large vector computation units.  \textit{Inhale} maximizes the inherent parallelism and bitline computational capability of memory subarrays to provide a high-throughput, low-latency, and energy-efficient solution for the SHA-3 algorithm with negligible area overhead (less than 2\% compared to conventional SRAM arrays). 
The benefits of \textit{Inhale} are enabled by three main reasons; (1) Based on our analysis, we find that 76\% of the operations in SHA-3 are shift operations. \textit{Inhale} reduces about 90\% of shift operations by proposing an effective \textit{Lane Per Row (LPR)} data alignment technique strategy to implicitly perform the shift operations using a memory-mapped controller. (2) We present an efficient in-place read/write strategy to perform the majority of the computation on the fly, thus, reducing the number of cycles and memory usage for intermediate data. (3) \textit{Inhale} design reuses the same memory array and peripherals to implement all the stages of SHA-3, thus significantly saving area. The combination of these benefits makes \textit{Inhale} yield significantly higher throughput-per-area-per-energy compared to the state-of-the-art.  

Unlike prior work, our design does not expand the TCB to the off-chip components (similar to Intel SGX \cite{costan_intel_2016} and AMD SME \cite{kaplan_amd_2016}).
Though SHA-3 is discussed here as a case-study, \textit{Inhale} is capable of computing other variants of the hash algorithm (e.g. SHA-1, SHA-2), block cipher algorithms (e.g. AES-256), as well as hash-based and finite-field-multiplication-based key establishment mechanisms and digital signature algorithms in the post-quantum era.

In summary, the paper makes the following \textbf{contributions:}

\vspace{-0.1cm}
\begin{itemize}
    \item We present \textit{Inhale}, an in-SRAM hashing engine with four-fold latency, throughput, area, and energy benefits for accelerating SHA-3 by exploiting compatible data alignment with minimal hardware modification. \textit{Inhale} can be realized either by re-purposing existing last-level caches or as a separate on-chip crypto-engine. Our proposed architecture provides the effective use of memory capacity with an efficient read/write strategy and flexible support for varying message sizes in hash computation, which can accommodate a broad range of SRAM array sizes on a wide variety of IoT devices.
    
    \item To reduce latency, we suggest a data alignment optimized for shift-intensive applications such as SHA-3. By implicitly performing \textit{inter-lane} shift operations via the controller, a significant number of shift operations in the SHA-3 are replaced by in-situ control commands, which not only reduces latency but also, in conjunction with an efficient read/write strategy, effectively reduces the capacity overhead of intermediate variables.
    
    \item We present two different variations of \textit{Inhale}, \textit{Inhale-Flex} and \textit{Inhale-Opt}, and evaluate them on SRAM and ReRAM-based memory arrays. The results demonstrate that on the same memory technology, \textit{Inhale} is able to achieve up to 172$\times$ throughput-per-area-per-energy compared to state-of-the-art in-memory implementations. 
    To separate the architectural and technology contributions of \textit{Inhale} and have an apples-to-apples comparison with prior work, we evaluate SHINE \cite{nagarajan_shine_2019} on SRAM (SHINE-SRAM) and find that our solution achieves 3.6$\times$ higher throughput-per-area than SHINE-SRAM due to architectural contribution and 3.4$\times$ better throughput-per-area due to the technology benefits.
\end{itemize}
\section{Background and Threat Model}

\subsection{SHA-3 Algorithm} \label{background-sha}
\begin{figure}[htp]
\centerline{\includegraphics[width=2.8in]{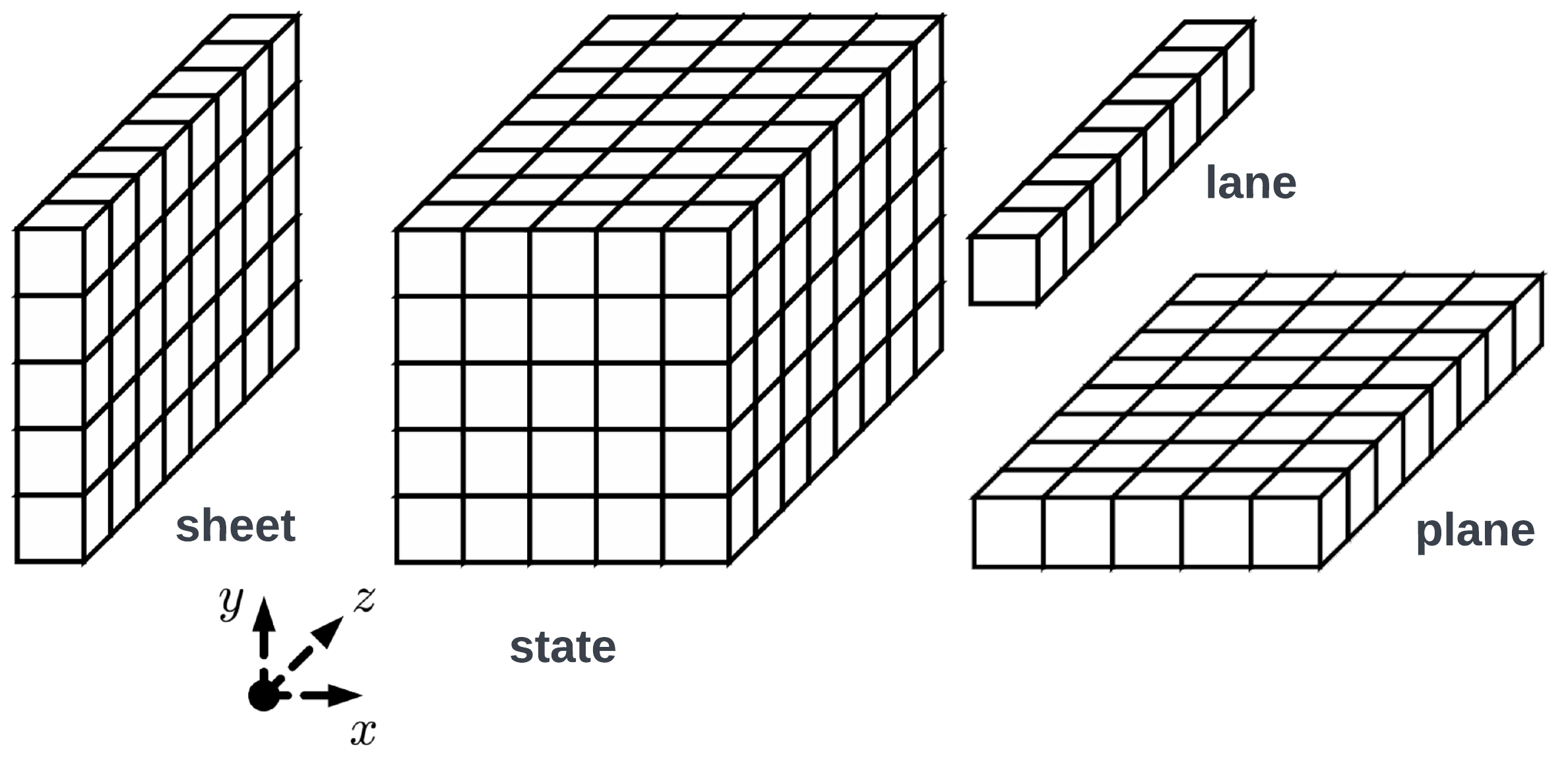}}
\caption{Naming conventions for parts of Keccak-$f$ \textit{State} \cite{dworkin_sha-3_2015}.}
\label{namingstate}
\end{figure}

The construction and verification of digital signatures, key derivation, and the generation of pseudo-random bits are the primary functions of the hash algorithm, which is an essential component of the information security domain.
SHA-3 is the third generation of standard hash functions, based on the implementation of the Keccak algorithm. The SHA-3 algorithm, unlike its predecessors, is a permutation-based cryptographic function.
SHA-3 hashing heavily relies on the structure of the Keccak sponge function.
The sponge structure is capable of data transformation, i.e., transforming arbitrary-length inputs into arbitrary-length outputs.
The fundamental function in Keccak is a permutation selected from the set of seven Keccak-$f$ permutations, abbreviated Keccak-$f[b]$, where $b\in\{25,50,100,200,400,800,1600\}$ is the width of the permutation.
In the sponge structure, the width of the permutation corresponds to the width of the \textit{State}.
The \textit{State} is organized as a grid of five-by-five \textit{Lanes} whose length is $w\in\{1,2,4,8,16,32,64\}$ and $b=25w$. 
The naming conventions for parts of the Keccak-$f$ \textit{State} used throughout the paper are shown in Fig. \ref{namingstate}.
\textit{In this paper, we discuss Keccak-f[1600] for SHA3-256.
However, \textit{Inhale} architecture can simply support all the Keccak permutations by adjusting the number of columns each \textit{Lane} occupies in the array thanks to our \textit{lane-per-row} data layout. }


Keccak has five stages of computing ($\theta,\rho,\pi,\chi,\iota$) in one round. Let's take one round of Keccak-$f[1600]$ as an example. Suppose we have a 1600-bit input \textit{State}, which according to the naming conventions (Fig. \ref{namingstate}), we can refer to as $A[x,y,z]$, where $x,y\in{0...4},z\in{0...63}$. The computations in five stages are as follows \cite{dworkin_sha-3_2015}:
\begin{enumerate}
    \item[$\theta$:] $C[x] = parity(A[x,0...4]), \hfill \forall x \in \{0...4\}$\\
                     $CT[x] = C[x] <<< 1, \hfill \forall x \in \{0...4\}$\\
                     $FT[x] = C[x-1] \oplus CT[x+1], \hfill \forall x \in \{0...4\}$
                     $A[x,y] = A[x,y] \oplus FT[x], \hfill \forall (x,y) \in \{0...4,0...4\}$
    \item[$\rho$:] $A[x,y] = A[x,y] <<< r(x,y), \hfill \forall (x,y) \in \{0...4,0...4\}$
    \item[$\pi$:]  $B[y,2x+3y]=A[x,y], \hfill \forall (x,y) \in \{0...4,0...4\}$
    \item[$\chi$:] $A[x,y]=B[x,y]\oplus (\neg B[x+1,y] \wedge B[x+2,y]),$\\
                    \hspace*{0pt}\hfill $\forall (x,y) \in \{0...4,0...4\}$
    \item[$\iota$:] $A[0,0]=A[0,0] \oplus RC[i]$,
\end{enumerate}
where the constant $r(x,y)$ in the $\pi$ stage is the rotation offset depending on the location of the \textit{Lane}, and $RC[i]$ in the $\iota$ stage is the round constant which is different for each of 24 rounds in Keccak-$f[1600]$. For different $w$, the number of rounds is equal to $12+2log_2w$. The output of each round will be XORed with the next piece of the message (1088 bits in Keccak-$f[1600]$) as the input of the next round of hashing \cite{dworkin_sha-3_2015} (see Fig. \ref{alignment}(c)).

\subsection{Computing in SRAM}
\begin{figure}[htp]
\centerline{\includegraphics[width=2in]{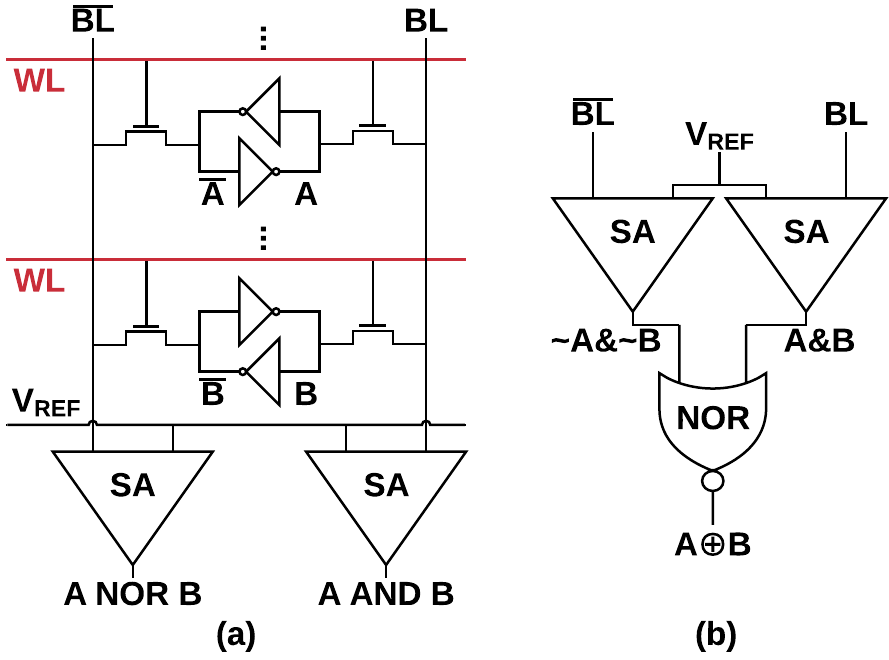}}
\caption{The XOR bitline operation with 6T SRAM cells. \cite{aga_compute_2017}}
\label{bitline}
\end{figure}

By activating more than one row in the SRAM subarray, in-SRAM processing can perform bitline computing \cite{jeloka_28_2016}. 
Fig. \ref{bitline} represents the logical operations that 6T SRAM can support. 
Using several activated wordlines and sensing amplifiers (SAs), AND and NOR operations are directly implemented in SRAM, as shown in Fig. \ref{bitline}(a). 
The SA on the bitline ($BL$) can only detect a voltage greater than $V_{ref}$ if every cell in the activated rows connected to the $BL$ contains the value `1'.
This indicates that the SA will detect the value `1', achieving element-wise AND operation.
The SA on the bitline-bar ($\overline{BL}$) will sense a voltage higher than $V_{ref}$ only if all the cells in the activated rows connected to the corresponding $\overline{BL}$ contain `1', which, in turn, implies that all the cells in the activated rows connected to the corresponding $BL$ contain `0'.  
Only if all the cells in the active rows connected to the corresponding bitline-bar ($\overline{BL}$) contain `1' would the SA on the $\overline{BL}$ detect a voltage greater than $V_{ref}$, which suggests that all the cells in the activated rows connected to the corresponding bitline contain `0'.
This indicates that the SA will detect the value `1', achieving element-wise NOR operation.
With the ability of logical bit-wise AND and NOR operations, the XOR operation can be achieved, as shown in Fig. \ref{bitline}(b).

Several studies based on SRAM computing are proposed \cite{aga_compute_2017,eckert_neural_2018,fujiki_duality_2019,subramaniyan_cache_2017}.
Cache Automaton employs a sense-amplifier cycling mechanism to read out several bits in a single time slot, hence lowering input symbol match time by a large margin \cite{subramaniyan_cache_2017}.
Compute Cache increases the logical operations by slightly changing the SA design, based on the specified NOR, AND, and XOR operations in \cite{aga_compute_2017}.
In this paper, we employ the XOR capability described in \cite{aga_compute_2017}.

\subsection{Threat Model}
Following the standard Trusted Execution Environment (TEE) threat model \cite{costan_intel_2016}, we assume that only the CPU itself is trusted and considered secure. The CPU contains no hardware backdoors. An attacker cannot directly or indirectly obtain access to the CPU's data.
\begin{figure*}[htp]
\centerline{\includegraphics[width=6.8in]{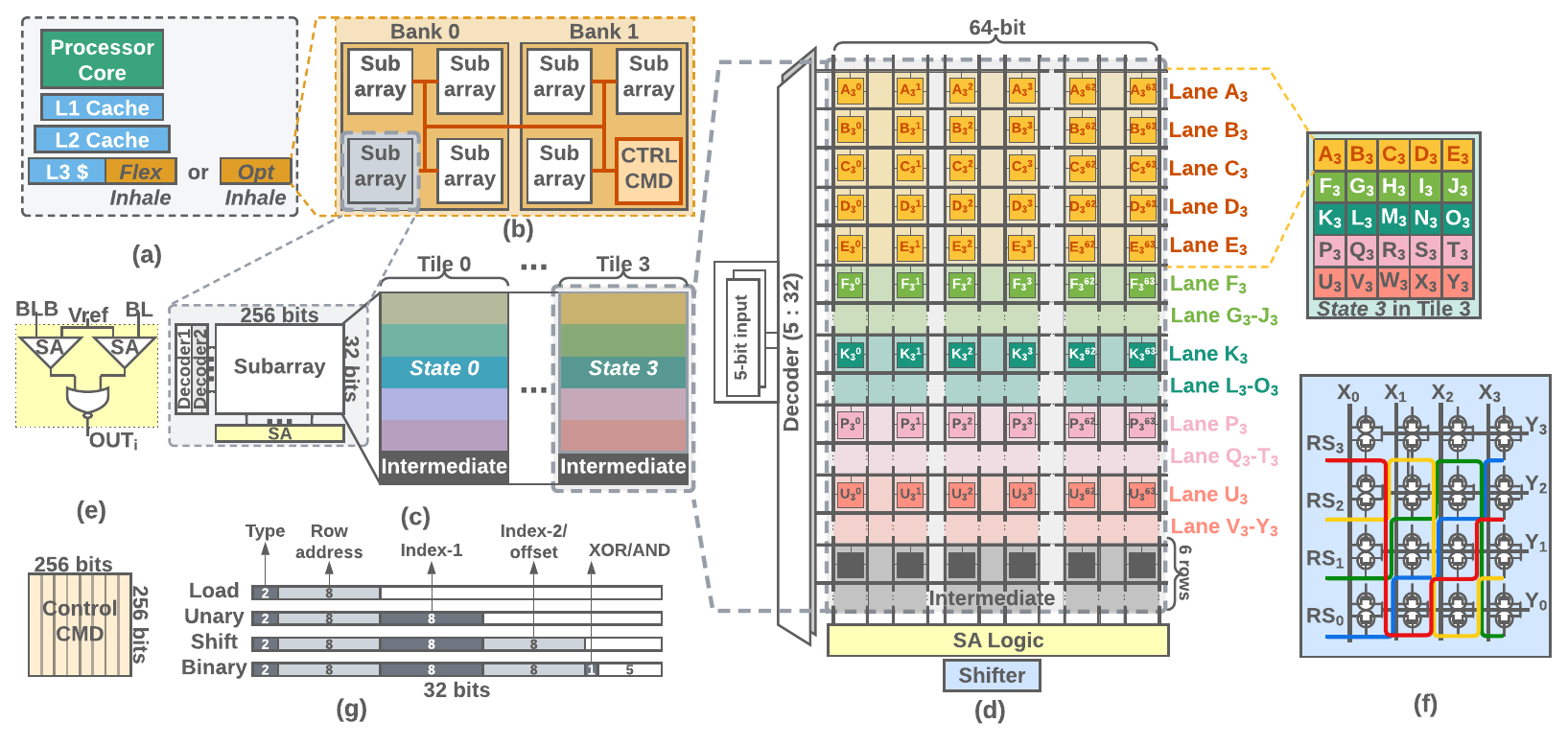}}
\caption{(a) \textit{Inhale} overview. (b) Two banks consisting of 8 subarrays, one of which stores control commands, working as an in-situ controller. (c) Data organization in one \textit{Inhale-Opt} subarray. (d) The detailed structure of the \textit{State} organization in \textit{Inhale-Opt} subarray. (e) The structure of the sense amplifier to support XOR operation. (f) The structure of the barrel shifter to support bidirectional rotate shifting. (g) Control signals for different operations in \textit{Inhale-Opt}.}
\label{overall}
\end{figure*} 

\vspace{-0.1cm}
\section{Related Work}
Several hardware acceleration solutions are proposed to improve the throughput, latency, or hardware overhead of Keccak/SHA-3. Tillich et al. \cite{tillich_high-speed_2009} use small LUTs to store intermediate variables to reduce the latency. Akin et al. \cite{akin_efficient_2010} and Wong et al. \cite{wong_new_2018} exploit a pipelined design to implement different stages of hashing algorithms to improve the throughput. Pessl et al. \cite{pessl_pushing_2013} implement SHA-3 on RFID by interleaving two adjacent \textit{Lanes} to generate a 128-bit word for computation in order to reduce the necessary on-chip memory space, and Ramanarayanan et al. \cite{ramanarayanan_18gbps_2010} fabricated a reconfigurable hashing accelerator for multi-mode SHA algorithms using unified SHA bit-slices and configurable compression circuits. However, none of them optimizes for high-throughput, low-latency, energy-efficiency, and area-efficiency all at the same time.


Moreover, recent in-memory solutions for SHA-3 implementation are proposed. Bhattacharjee et al. proposed an implementation of SHA-3 based on ReRAM \cite{bhattacharjee_sha-3_2017}. Twenty-five \textit{Lanes} are aligned as a grid of 5 by 5 \textit{Lanes}, and the 5-input XOR is used to provide high parallelism in computation. Although the authors use techniques such as stage fusion to reduce computational overhead, their implementation of basic XOR requires 12 cycles, thus significantly increasing the overall computational latency. 
Nagarajan et al. present a novel ReRAM-based SHA-3 accelerated solution, named \textit{SHINE} \cite{nagarajan_shine_2019}. The authors construct multi-bit XOR, 64-bit and arbitrary bit shift operations based on ReRAM arrays and employ these cascaded arrays for pipeline processing to achieve a high throughput SHA-3 solution. To optimize for latency, the authors present a pipeline design by replicating the memory arrays in each stage of the pipeline, thus trading hardware overhead and throughput-per-unit-area for latency. Moreover, their design does not generalize to other hashing algorithms. In addition, it expands the trusted computing base (TCB) to the off-chip memory which exposes more space to attackers.
Zhang et al. present \textit{Recryptor}, a general-purpose processor with in-memory computing capabilities to handle a variety of cryptographic algorithms for IoT security \cite{zhang_recryptor_2018}. To implement \textit{inner-lane} and \textit{inter-lane} shifts for $\rho$ and $\pi$ stages, \textit{Recryptor} requires a large bit width or a multi-stage shifter, which can result in a high hardware overhead. 
Moreover, due to the complex peripheral circuitry, Recryptor only runs at 28.8 MHz, which significantly limits the performance.


While it is possible to improve individual metrics, to efficiently enable secure edge computing on resource-constrained IoT devices, a four-fold optimization for latency, throughput, area, and energy efficiency is required.
In Section \ref{sec:evaluation}, we compare the performance, area and energy consumption of \textit{Inhale} with \cite{tillich_high-speed_2009,pessl_pushing_2013, akin_efficient_2010, wong_new_2018,nagarajan_shine_2019,  zhang_recryptor_2018}.

\section{Implementation}

\subsection{Overview}
This section introduces \textit{Inhale}, our proposed in-SRAM hashing architecture. \textit{Inhale} can be realized either by re-purposing existing last-level caches in the device, named \textit{Inhale-Flex}, or by developing a specialized on-chip SRAM-based acceleration engine, named \textit{Inhale-Opt}, as depicted in Fig. \ref{overall} (a-b). \textit{Inhale-Opt} trade flexibility for latency, throughput, area, and energy benefits by reducing the subarrays to size 32$\times$256. This reduces the subarray size and read/write latency, which enables higher frequencies and higher throughput.


\textit{Inhale}'s smallest working unit requires two subarrays, one for data storage and computation and the other one for control and command data. 
Fig. \ref{overall}(b) represents two banks, each with four subarrays, to illustrate our design. Based on the application requirements and power budget, $b$ banks can be utilized to hash up to $b\times (n-1) \times \left \lceil\frac{c}{w} \right \rceil$ independent messages in parallel, where $n$ is the number of subarrays in two adjacent banks, $c$ is the number of columns in each subarray, and $w$ is the width of a \textit{State} (see Section \ref{background-sha}). 
The rest of this section discusses data organization, implementation of each stage, the memory-mapped controller, and the generality of \textit{Inhale}.

\begin{figure*}[htp]
\centerline{\includegraphics[width=6.4in]{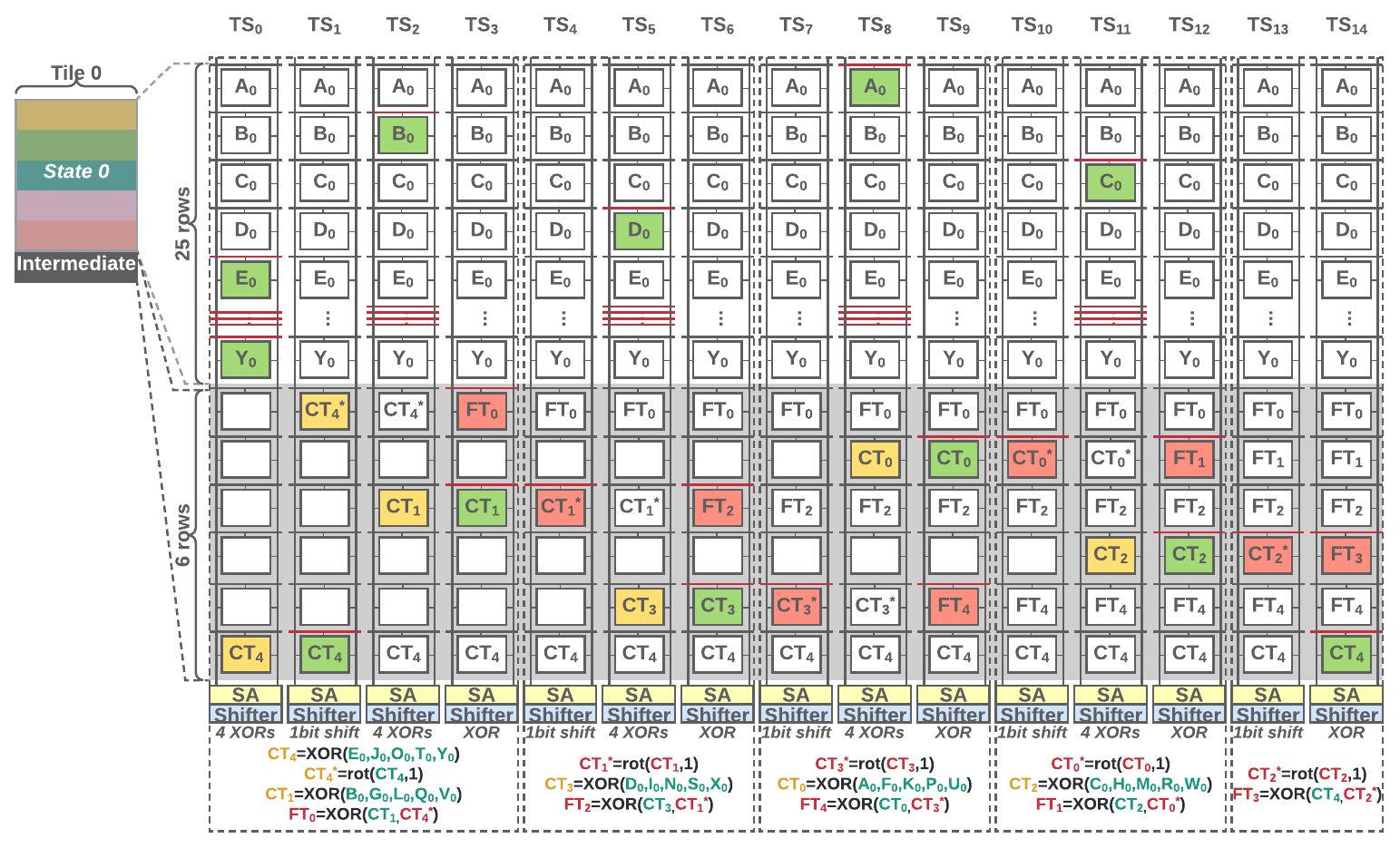}}
\caption{Explaining in-place read/write strategy with only 6 intermediate rows in $\theta$ stage in several time steps (TS).}
\label{readwrite}
\end{figure*} 

\subsection{Data Organization} \label{dataorganization}

To efficiently utilize the limited SRAM capacity and to substantially reduce the number of shift cycles in $\pi$ and $\chi$ stages, the input and intermediate data of the SHA-3 algorithm are delicately placed in the same subarray, as shown in Fig. \ref{overall}(c)-(d). The size of the subarray in \textit{Inhale-Opt} design is 32$\times$256, which consists of 4 Tiles. Each Tile contains one \textit{State}, i.e., the inputs of the SHA-3 algorithm. 
In each Tile, the first 25 rows are used to store twenty five \textit{64-bit Lanes}, as we call it \textit{lane-per-row} data alignment. The rest of the rows are used for intermediate data. Since we use an in-place data read/write strategy, only 6 rows are enough to store intermediate data.

In \textit{Inhale-Flex} design (which repurposes existing cache arrays), the size of the SRAM subarray is assumed to be 256$\times$256 following ARM Cortex-M0+ microcontroller \cite{noauthor_lpc1225fbd48arm_nodate}. 
However, \textit{Inhale} design can work with any subarray with a minimum of 32 rows (to store 25 \textit{Lanes} and 6 intermediate data) and a minimum of 64 columns (to store a \textit{Lane} of 64-bit). 
If the row number of the array exceeds 32, the extra rows can be utilized to store longer messages or more states in the Tile.

Prior work \cite{zhang_recryptor_2018} arranges every 5 \textit{Lanes} (320 bits) in one row and require 5 rows to store an entire \textit{State}. 
Although this data alignment can highly expose the highest bit parallelism in $\theta$ stage, it incurs a significant number of shifts in $\rho$, $\pi$, and $\chi$ stages, where shift operations dominate. To reduce the shift cycles, they heavily rely on the high-overhead shifting modules, such as barrel shifter. 
\textit{Inhale} utilizes \textit{lane-per-row} data alignment, allowing all \textit{inter-lane} shifts to be implicitly implemented by the controller simply by selecting the intended \textit{Lane} instead of shifting. 

In summary, our proposed \textit{lane-per-row} data alignment has two advantages: 1) it significantly reduces the shift operations, which greatly reduces the need for large n-bit shifters. 
Almost every stage in the SHA-3 algorithm has \textit{inter-lane} shifts. 
Specifically, $\theta$, $\pi$ and $\chi$ requires \textit{inter-lane} shifts to perform data alignment before XOR operations (e.g. C[(\textit{x}-1)], C[(\textit{x}+1)] in $\theta$, A[(\textit{x}+3\textit{y})] in $\pi$, and A[(\textit{x}+1)], A[(\textit{x}+2)] in $\chi$ according to \cite{dworkin_sha-3_2015}.
\textit{Inhale} implements these shift operations implicitly, thus, introducing area, latency, and energy benefits, and 2) it enables flexibility of mapping computation on smaller subarray sizes (i.e., only 32 rows and 64 columns are enough for \textit{Inhale} to work), which are most efficient on resource-constrained edge devices. This is unlike prior work that requires the number of columns to be at least 320.
Although we are discussing the data alignment for Keccak-$f[1600]$ in this paper (the largest setting), our data alignment can also accommodate other settings of Keccak by simply adjusting the number of bits per \textit{Lane}.




\subsection{Implementation of SHA-3 Stages}

\textbf{Theta ($\theta$) Stage:}
Theta ($\theta$) stage includes a series of XOR, \textit{inter-lane} shift, and \textit{intra-lane} shift operations on the original \textit{State} of 1600 bits. In general, all \textit{Lanes} within the same \textit{Sheet} (Fig. \ref{namingstate}) need to perform bit-wise XOR operations to obtain an additional intermediate \textit{Plane}. Then, each \textit{Lane} is bit-wise XORed with its corresponding \textit{Lane} in the intermediate \textit{Plane} to obtain the final \textit{Plane}. At the last step, each \textit{Lane} is bit-wise XORed with the corresponding \textit{Lane} in the final \textit{Plane}. The bit-wise XOR operation here is achieved by using two decoders to activate two rows simultaneously. We follow the bitline XOR functionality implementation in \cite{aga_compute_2017}, which costs 3 times longer than a single subarray read/write access.

Fig. \ref{readwrite} demonstrates how the in-place read/write strategy is implemented in several time steps (TS) and explains why only 6 intermediate rows are enough for 1 round of computation. The first 25 rows store the 25 \textit{Lanes}, and the next 6 rows store the intermediate data. 
In the first 4 time steps ($TS_0-TS_3$), 4 XORs for $CT_4$, one 1-bit shift, 4 XORs for $CT_1$, and a single XOR are done. In particular, in $TS_3$, the first in-place operation is performed. The $CT_4^*$ generated by $TS_2$, is activated with the line where $CT_1$ is located, and then, the XOR operation is performed. The result $FT_0$ is written back to $CT_4^*$, i.e., \textit{in-place computing}. All the cells with the in-place operations are marked in red. In $TS_4-TS_{14}$, 8 time steps utilize the in-place operations. Thus, we are able to save 40\% space for the intermediate rows. This, in turn, results in the fewer number of required rows in \textit{Inhale} design, and effectively translates to shorter read/write latency.

After we generate the five \textit{Lanes} in the intermediate rows ($FT_0-FT_4$), we need to XOR the different \textit{Lanes} located in the same \textit{Sheet} with the corresponding \textit{Lane} in the intermediate rows. For example, $FT_0$ needs to be respectively XORed with $A_0$, $F_0$, $K_0$, $P_0$, and $U_0$, and overwrite the \textit{Lanes} in the \textit{State} with the obtained results. The complete correspondences are $FT_0-(A_0,F_0,K_0,P_0,U_0)$, $FT_1-(B_0,G_0,L_0,Q_0,V_0)$, $FT_2-(C_0,H_0,M_0,R_0,W_0)$, $FT_3-(D_0,I_0,N_0,S_0,X_0)$, $FT_4-(E_0,J_0,O_0,T_0,Y_0)$.
Theta stage takes a total of 210 cycles, comprising 50 XOR operations and 5 \textit{inner-lane} 1-bit shift operations for 25 \textit{Lanes}.

\textbf{Rho ($\rho$) and Pi ($\pi$) Stage:}
For the Rho stage, all the \textit{Lanes} need to be \textit{inner-lane} shifted with a certain number of bits according to their positions in the \textit{State}. For example, the $C_0$ at $(2,0)$ needs to be shifted 62 bits to the right. Due to the randomness characteristic of the shift in the Rho stage, we use a 64-bit barrel shifter to perform these irregular shifts. 
For the Pi stage, since we use the \textit{lane-per-row} data organization, the \textit{inter-lane} shift in the Pi stage can be implicitly implemented by the memory-mapped controller. Specifically, no physical operation is done in the Pi stage. However, the controller selects the intended row (i.e., \textit{Lane}) later in the Chi stage to get the correct data.
Rho stage incurs a total of 50 cycles for 25 \textit{inner-lane} shift operations, whereas the Pi stage is implicitly completed by an in-situ controller at no cost.

\textbf{Chi ($\chi$) and Iota ($iota$) Stage:}
In the Chi stage, each \textit{Lane} needs to perform a series of operations with the next two \textit{Lanes} that are on the same \textit{Plane}. Since all its operations depend only on the data in the same \textit{Plane}, we can use the in-place read/write strategy (similar to the Theta stage) to perform operations \textit{Plane}-by-\textit{Plane} to save intermediate rows.
Specifically, to generate the results of the five \textit{Lanes} (e.g., $A,B,C,D,E$) in the first \textit{Plane} after the Chi stage, we first use bitline computing to perform NOT operations on each of the five \textit{Lanes}, and write the results of NOTs ($\neg A, \neg B, \neg C, \neg D, \neg E,$) into the five intermediate rows. Then we perform AND operations between these \textit{Lanes} and corresponding original \textit{Lanes} (e.g., $\neg A \wedge B, \neg B \wedge C, \neg C \wedge D, \neg D \wedge E, \neg E \wedge A$) and write the results back to the corresponding row in the intermediate row. Finally, the original \textit{Lanes}, which have not been overwritten from the beginning, are now XORed with the corresponding \textit{Lane} in the intermediate rows, and the results of XORs are written back to the original position instead of the intermediate rows. Since we only compute the next \textit{Plane} after finishing the computation of the previous \textit{Plane}, only 5 intermediate rows are required.

In the Iota stage, we only need to bit-wise XOR the first \textit{Lane} with pre-computed 64-bit data ($RC[n]$). Since the data contains only 64 bits, we can use the control signal consisting of the data for data transfer without storing twenty-five 64-bit data in the subarray in advance. By placing this data in different Tiles, we can implement the XOR operation in the Iota stage in parallel.
When three consecutive \textit{Lanes} require three XORs and two implicit \textit{inter-lane} shift operations, the Chi stage's 75 XOR operations require a total of 300 cycles. Iota is a trivial stage, requiring only four cycles for one XOR and a write-back in one round.

\subsection{Control Implementation}
The control of the \textit{inhale} is implemented by an inter-bank controller and a control subarray storing pre-generated control signals.
As depicted in Fig. \ref{overall}(b), the inter-bank controller is responsible for retrieving control signals from the control subarray and broadcasting them to the target subarrays during the corresponding clock cycles.
The pre-generated control commands are stored in the control subarray in the format depicted in Fig. \ref{overall}(g).
The first two bits (i.e., [0:1]) specify one of the four commands: LOAD command for loading Round Constants (RCs), UNARY command for NOT operation, SHIFT command for shift operation, and BINARY command for XOR and AND operation.
For subarrays with different numbers of rows, the row indexes are represented with different bit widths.
In 256$\times$256 subarrays, [2:9] provides the row index of the command's result.
The LOAD instruction contains the row index of the row to be written. Then, the 64-bit RC is created and delivered to the target subarrays. 
After receiving the RC, the target subarray will write them to the intermediate rows of different Tiles in order to perform the computation.
[10:17] in the UNARY command represents the row index of its operand.
BINARY and SHIFT have two operands. Therefore, bits in the location of [18:25] specify the row index of the second operand (or the shift offset).
The 26-th bit of BINARY command indicates, by a value of 0 or 1, whether the operation is XOR or AND.
Longer control commands can be employed to handle larger subarrays.
The memory-mapped controller broadcasts control signals to all arrays in the two adjacent banks, and the timely arrival of the control signals can be ensured by the pre-fetching techniques.

\begin{table*}
	\centering 
	\caption{Comparison of different designs.}
	\scalebox{0.9}{
	\begin{tabular}{>{\centering\arraybackslash}p{1.7in}>{\centering\arraybackslash}p{0.4in}>{\centering\arraybackslash}p{0.4in}>{\centering\arraybackslash}p{0.35in}>{\centering\arraybackslash}p{0.45in}>{\centering\arraybackslash}p{0.35in}>{\centering\arraybackslash}p{0.35in}>{\centering\arraybackslash}p{0.65in}>{\centering\arraybackslash}p{0.35in}>{\centering\arraybackslash}p{0.9in}}
		\toprule \midrule
		&  Tech. &  Max $f$ (MHz) &  Area (KGE) &  Latency (cycles) &  Latency (ns) &  Tput. (Mbps) &  Tput./Area (Mbps/KGE) &  Energy (nJ) & Tput./Area/En. (Mbps/(KGE$\cdot$nJ))\\ \midrule
		\textit{Inhale-Opt-SRAM} & 28nm & 6.7K & \textbf{63.6} & 564 & 83.6 & 52K &  \textbf{818} & \textbf{0.456} & \textbf{1.8K}\\ 
        \textit{Inhale-Flex-SRAM} & 28nm & 6.1K & 386 & 564 & 91.9 & 47.3K &  123 & 0.596 & 206\\ \addlinespace
        SHINE-1-SRAM & 28nm & 6.7K & 494 & 264 & 39.1 & 111K &  225 & - & -\\ 
        SHINE-2-SRAM & 28nm & 6.7K & 717 & 140 & \textbf{20.7} & \textbf{210K} &  293 & - & -\\ \addlinespace
        Recryptor\cite{zhang_recryptor_2018} & 40nm & 28.8 & 600 & 139 & 4.8K & 226 &  0.377 & 2.03 & 0.186\\ \midrule
        \textit{Inhale-Opt-ReRAM} & 28nm & 2.4K & \textbf{19.1} & 564 & 235 & 18.6K &  \textbf{970} & \textbf{0.348} & \textbf{2.79K} \\
        \textit{Inhale-Flex-ReRAM} & 28nm & 2.3K & 56.3 & 564 & 240 & 18.1K &  322 & 0.446 & 721\\  \addlinespace
        SHINE-1-ReRAM\cite{nagarajan_shine_2019} & 65nm & 2K & 494 & 264 & 132 & 33K &  66.8 & 4.13 & 16.2\\ 
        SHINE-2-ReRAM\cite{nagarajan_shine_2019} & 65nm & 2K & 717 & 140 & 70 & 62.2K &  86.7 & 3.5 & 24.8\\ 
        SHINE-1-ReRAM (projected) & 28nm & 4.6K & 494 & 264 & 56.9 & 76.5K &  155 & - & -\\ 
        SHINE-2-ReRAM (projected) & 28nm & 4.6K & 717 & 140 & \textbf{30.2} & \textbf{144K} &  201 & - & -\\  \midrule
        Akın\cite{akin_efficient_2010} & 90nm & 455 & 10.5K & 25 & 54.9 & 19.8K &  1.89 & \textgreater43.5 & $<$0.043\\ 
        Tillich\cite{tillich_high-speed_2009} & 180nm & 488 & 56.3K & 25 & 51.2 & 21.2K &  0.377 & \textgreater43.5 & $<$0.009\\ 
        Pessl-V1 \cite{pessl_pushing_2013} & 130nm & 1 & 5.5K & 10.7K & 10.7M & 0.102 &  18.5e-6 & \textgreater43.5 & $<$4.25E-7\\ 
        Pessl-V2 \cite{pessl_pushing_2013} & 130nm & 1 & 5.9K & 7.4K & 7.4M & 0.147 &  24.9e-6 & \textgreater43.5 & $<$5.73E-7\\ 
        Wong\cite{wong_new_2018} & 65nm & 1K & 105K & -& -& 48K &  0.457 & \textgreater43.5 & $<$0.011\\ \midrule
		\bottomrule
		\addlinespace
	\end{tabular}}
	\label{resulttable}
\raggedright\small \\ \emph{*} Energy of SHINE variants could not be simulated due to a lack of specifications, represented by "-" in the Table. \\
\emph{$\dagger$} SHINE-1 (optimized for area and power efficiency) and SHINE-2 (optimized for throughput and latency) are two variants of ReRAM-based SHA-3 accelerator presented in \cite{nagarajan_shine_2019}.
\end{table*}

\subsection{Generality of \textit{Inhale} Architecture}

\begin{figure}[htp]
\centerline{\includegraphics[width=2.75in]{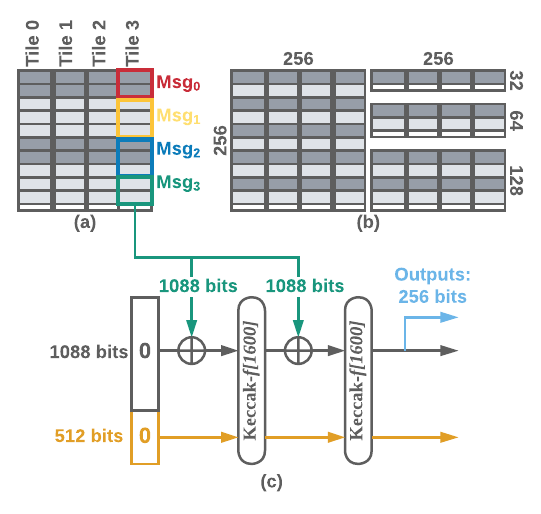}}
\caption{(a) \textit{Inhale} supports different lengths of messages. (b) Different subarray sizes and data organizations compatible with \textit{Inhale}. (c) Message split in Keccak-$f[1600]$.}
\label{alignment}
\end{figure}

With \textit{lane-per-row} data alignment, \textit{Inhale} can accommodate messages of varying lengths (i.e., M bit, where M can be smaller or larger than 1088). In SHA-3, the hash value of a message with fewer than 1088 bits can be generated from 25 \textit{Lanes}. For a message longer than 1088 bits, $(M$ $mod$ $1088) + 8$ \textit{Lanes} are required to calculate its hash value. For example, in Fig. \ref{alignment}(a), the last Tile has four independent messages (Msg$ _0$, ..., Msg$ _3$). Assume that Msg$ _0$ has 2168 bits. Using Keccak$f[1600]$ as an example, Msg$_0$ is split into two pieces, each of which consists of 1088 bits (i.e., requires seventeen 64-bit \textit{Lanes}). Including the 8 \textit{Lanes} shared by the two components, Msg$ _0$ should occupy a total of 42 rows.
Unlike prior work that uses \textit{plane-per-row} data alignment (i.e., storing the data of one \textit{Plane} in a row), which cannot reuse the shared 8 \textit{Lanes}, \textit{Inhale} reuses the shared \textit{Lanes} without modifying the data alignment or incurring any additional computational expense, hence reducing latency and capacity overhead.

\textit{Inhale} can also support subarrays of different sizes. As shown in Fig. \ref{alignment}(b), a 256$\times$256 subarray can be used to hold four Tiles, each of which can compute up to ten independent messages. Alternatively, we can use a 32$\times$256 subarray to store four independent messages to enable parallel processing. \textit{Inhale}'s extensive support for variable subarray sizes enables its deployments on diverse IoT devices with varying SRAM array sizes. Note that the performance of different subarray sizes depends on read/write latency. In the evaluation section, we evaluate and compare the performance and overhead of the common 256$\times$256 subarray size (i.e., \textit{Inhale-Flex}) and the smallest subarray size of 32$\times$256 (i.e., \textit{Inhale-Opt}).

\section{Evaluation}\label{sec:evaluation}

\subsection{Evaluation Methodology}
In this section, we evaluate the latency, area, throughput, energy, throughput-per-area-per-energy, and performance scaling of \textit{Inhale-Flex} and \textit{Inhale-Opt}, 
and compare them to prior in-memory solutions \cite{zhang_recryptor_2018,nagarajan_shine_2019} and other ASIC implementations \cite{pessl_pushing_2013,wong_new_2018,tillich_high-speed_2009,akin_efficient_2010}, as shown in Table \ref{resulttable}.
We use DESTINY simulator for area and energy simulation \cite{poremba_destiny_2015}. 
To decouple architectural contribution from the technology contribution and have an apples-to-apples comparison between SHINE and \textit{Inhale}, we evaluate \textit{Inhale} in 28nm ReRAM and SRAM, and also evaluate SHINE in 28nm SRAM and ReRAM.
To decouple the area overhead from the technology node, we normalize the area overhead to the number of Kilo Gate Equivalent (KGE) (i.e., the area of a single NAND gate in the corresponding technology node).


The read and write access latency of 6T SRAM subarrays are extracted from the Synopsys Design Compiler with 45nm process and then projected to 28nm process. PyMTL3 \cite{jiang_pymtl3_2020} and OpenRAM \cite{guthaus_openram_2016} are used to generate SRAM subarrays.
The read and write latency of the ReRAM array are calculated from DESTINY simulator.
The latency number in Table \ref{resulttable} represents the time consumed by one round of SHA-3 for each of the designs.

\subsection{Latency Comparison}


While \textit{Inhale-Flex} design is based on 256-row memory arrays, \textit{Inhale-Opt} design utilizes 32-row memory arrays, thus, having lower latency (10\% reduction in conservative conditions). This is mainly because shorter subarrays have shorter bitlines, which results in lower read/write latency and higher frequency. 
Moreover, \textit{Inhale-Opt-SRAM} and \textit{Inhale-Flex-SRAM} designs achieve 58$\times$ and 52$\times$ better performance compared to Recryptor \cite{zhang_recryptor_2018}, respectively. This is because Recryptor is constrained by its complicated peripheral hardware (two-stage shifter and many MUXs) and can only operate at 28.8 MHz, resulting in a considerable performance limitation.

The SHINE-SRAM in 28nm technology has 53\%-75\% lower latency compared to the \textit{Inhale-SRAM} design in the same technology node. 
This is mainly because SHINE uses a dedicated 5-bit XOR and 3-bit XOR logic that can compute 5-bit XOR or 3-bit XOR in four clock cycles. However, \textit{Inhale} compute 5-bit XOR by performing in-SRAM XOR functionality for 2 bits at a time, each taking 3 clock cycles (i.e., ((($b_0$ $\oplus$ $b_1$)$\oplus$ $b_2$) $\oplus$ $b_3$) $\oplus$ $b_4$), thus, a 5-bit XOR taking 12 cycles, a 3-bit XOR taking 6 cycles, and a 2-bit XOR requires three clock cycles. 
\textit{In summary, SHINE design trade the area-overhead to achieve 53\%-75\% lower latency compare to \textit{Inhale}. However, the \textit{Inhale-SRAM} design incurs 7.7$\times$ - 11.2$\times$ lower area-overhead compared to the SHINE-SRAM in the same technology node by sacrificing only 53\%-75\% latency. As a result, \textit{Inhale-Opt-SRAM} can achieve 2.8$\times$ - 3.6$\times$ higher throughput per unit area than the SHINE-SRAM design, and \textit{Inhale-ReRAM} can achieve 4.8$\times$ - 6.3$\times$ higher throughput-per-unit-area than the SHINE-ReRAM design!
In addition, SHINE uses a large number of small-sized arrays and non-generalized interconnections, which sacrifices generality and introduces a large amount of peripheral circuitry.
}


In general, \textit{Inhale-SRAM} has a shorter latency than \textit{Inhale-ReRAM}. This is due to the fact that the read and write latency of ReRAM are imbalanced, with the write latency being way larger than the read latency, resulting in a higher write latency than SRAM but having a lower read latency than SRAM. Since the maximum supported frequency is governed by the maximum value of the read and write latencies, \textit{Inhale-ReRAM} has a lower frequency than \textit{Inhale-SRAM}. 

\subsection{Area Overhead}



\begin{table}
	\centering
	\caption{Peripheral comparison in \textit{Inhale} vs \textit{SHINE}.}
	\scalebox{0.8}{
	\begin{tabular}{>{\centering\arraybackslash}p{0.35in}ccccc}
		\toprule
        & & SHINE-1 & SHINE-2 & \textit{Inhale-Flex} & \textit{Inhale-Opt}\\
		\midrule
		\#Cells && 407552 & 576000 & 65536 & 8192 \\
		\#Decoders && 178 & 261 & 2 & 2 \\
		\#SAs && 11392 & 16704 & 256 &  256\\
		\bottomrule
	\end{tabular}}
	\label{peripheral}
\end{table}

\textit{Inhale-Opt-SRAM} has 84\% lower area overhead compared to \textit{Inhale-Flex-SRAM}, and \textit{Inhale-Opt-ReRAM} has 66\% lower area overhead compared to \textit{Inhale-Flex-SRAM}. The difference in reduction ratio between SRAM and ReRAM-based implementation is mainly because the ratio of the memory array area to the peripheral circuitry area in ReRAM is lower than in SRAM.  
\textit{Inhale-SRAM} has 36\% - 89\% lower area overhead compared to Recryptor.
This is due to the fact that, unlike Recryptor, which requires a wide bit-width (320 bits) or a multi-stage shifter and MUXs, \textit{Inhale-SRAM} only requires a 64-bit shifter, hence minimizing the area overhead.

\textit{Inhale-Opt-SRAM} has 87\% and 91\% lower area overhead than SHINE-1-SRAM and SHINE-2-SRAM, respectively. 
This is mainly because SHINE employs more memory arrays and peripheral circuitry to implement the pipeline design. As shown in Table \ref{peripheral}, SHINE requires up to 70$\times$ more memory cells, up to 131$\times$ more decoders, and up to 65$\times$ more SAs than \textit{Inhale}.
Moreover, \textit{Inhale-ReRAM} has a 70\%-85\% lower area overhead than \textit{Inhale-SRAM}, mainly due to the density advantage of ReRAM arrays.

\subsection{Throughput per Area per Energy}


\textit{Inhale} design enables a high-throughput, low-latency, design energy-efficient with low hardware overhead, thus, yielding the highest throughput-per-area-per-energy (TAE) among all other designs. 
The throughput is defined by dividing the processed message bits per round (1088 bits in SHA3-256 \cite{dworkin_sha-3_2015}) by one round latency (Table \ref{resulttable}) and then multiplying the parallelism factor (4 in our evaluation).
\textit{Inhale-Opt} has 4$\times$ - 9$\times$ higher TAE than \textit{Inhale-Flex} due to the smaller array size (which also translates to lower read/write latency and lower energy consumption). 
\textit{Inhale-Opt-ReRAM} has 173$\times$ and 113$\times$ higher TAE than SHINE-1-ReRAM and SHINE-2-ReRAM, respectively. 
As discussed earlier, \textit{Inhale} uses significantly fewer memory cells and peripheral circuitry than SHINE, thus, achieving great area and energy benefits.
According to Table \ref{resulttable}, the TAE of \textit{Inhale-Opt-SRAM} is comparable to \textit{Inhale-Opt-ReRAM}. However, the TAE of \textit{Inhale-Flex-ReRAM} is 3.5$\times$ higher than \textit{Inhale-Flex-SRAM}. This is mainly because \textit{Inhale-Flex-SRAM} has 6.9$\times$ more area overhead than \textit{Inhale-Flex-ReRAM}, but only has 2.6$\times$ higher throughput. However, \textit{Inhale-Opt-SRAM} has 3.3$\times$ more area overhead than \textit{Inhale-Opt-ReRAM} with 2.8$\times$ higher throughput.

\subsection{Performance Scaling}

\begin{figure}[htp]

\subfloat[With Power Limitation]{%
  \includegraphics[clip,width=3.4in]{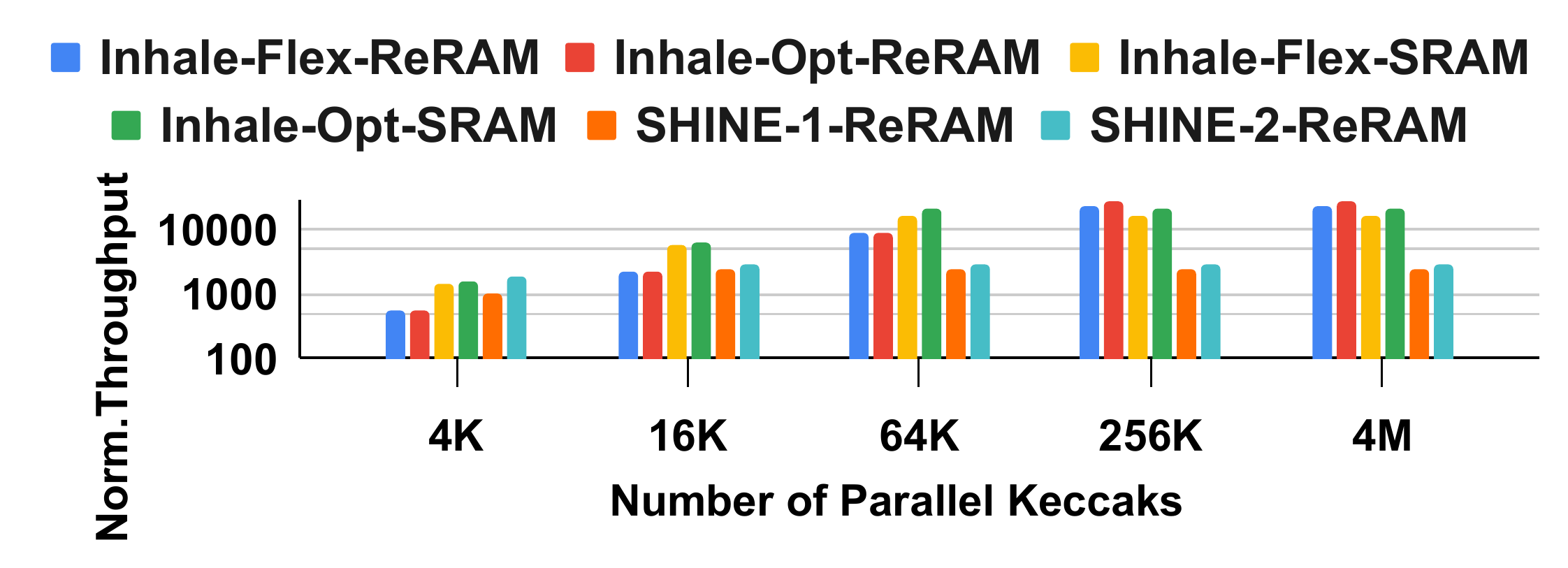}%
}

\subfloat[Without Power Limitation]{%
  \includegraphics[clip,width=3.4in]{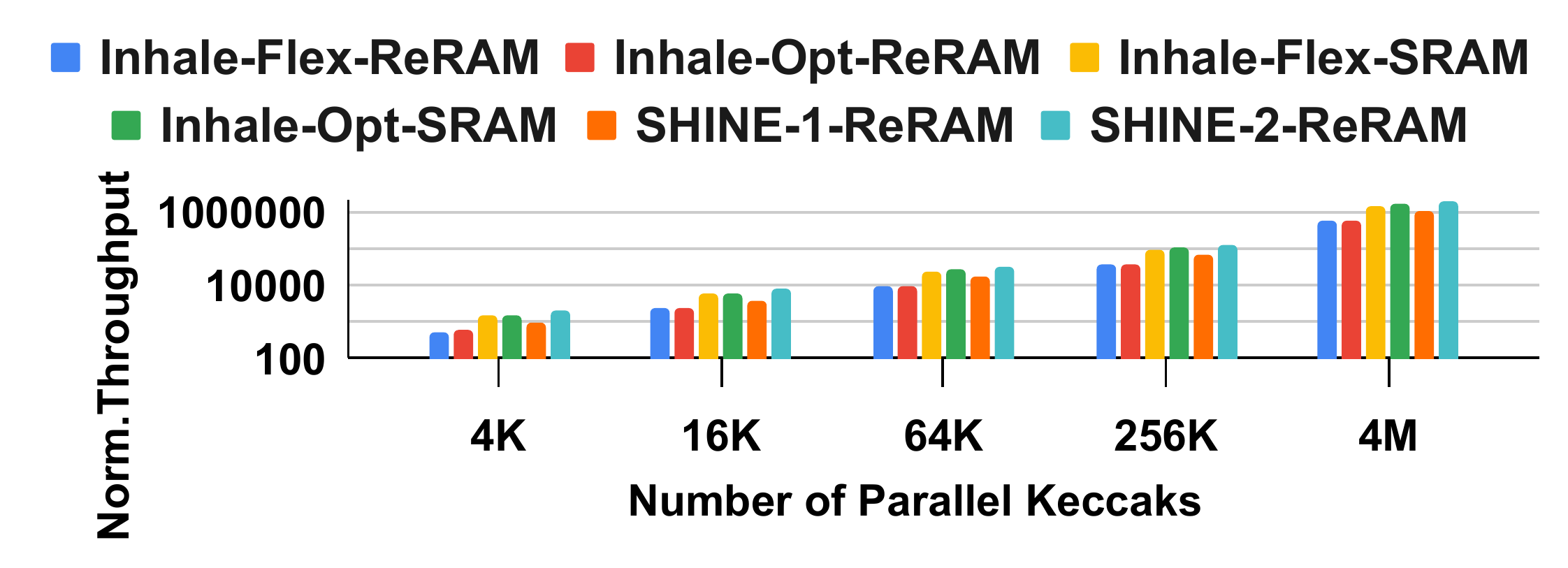}%
}

\caption{Performance scaling with the number of parallel Keccaks.}
\label{scaling}
\end{figure}

In this section, we study the scalability of different designs. To show the effect of more Keccak permutations on performance, we increase the number of parallel Keccaks to up to 4 million and evaluate the performance of \textit{Inhale} and SHINE with and without power limitations.
Fig. \ref{scaling} depicts the throughput of different \textit{Inhale} and \textit{SHINE} designs, normalized to the throughput of the single SHINE-1, with and without power constraints. In the power-constrained scenario, we assume the maximum power of 75W for all the designs. This reduces the number of computational units that can be executed concurrently.

Without taking the power constraint into account, we assume that all the parallel Keccaks can operate in processing units of \textit{Inhale} and SHINE in parallel without compromising the performance. In this scenario, the relationship between the performance of various designs is identical to that shown in Table \ref{resulttable}, i.e., the throughput is proportional to the number of parallel Keccaks. 
In the power-constrained scenario, \textit{Inhale-Opt-SRAM} has 2.9$\times$, 2.8$\times$, 1.1$\times$, 1.6$\times$, and 0.8$\times$ higher throughput than \textit{Inhale-Flex-ReRAM}, \textit{Inhale-Opt-ReRAM}, \textit{Inhale-Flex-SRAM}, SHINE-1-ReRAM, and SHINE-2-ReRAM, respectively, when computing 4K Keccak permutations in parallel.
However, when the number of parallel Keccaks reaches 4M, \textit{Inhale-Opt-SRAM} cannot continue the linear increase of throughput due to the power constraint, while \textit{Inhale-Opt-ReRAM} becomes the design with the highest throughput, and shows up to 1.3$\times$, 1.7$\times$, 1.3$\times$, 12$\times$, and 10$\times$ better throughput over \textit{Inhale-Flex-ReRAM}, \textit{Inhale-Flex-SRAM}, \textit{Inhale-Opt-SRAM}, SHINE-1-ReRAM, and SHINE-2-ReRAM, due to its lowest power consumption per permutation.

\section{Conclusion}

In this paper, we propose \textit{Inhale}, an in-SRAM hashing engine with four-fold latency, throughput, area, and energy benefits for accelerating SHA-3 on a CPU chip with narrowed trusted computing base. By leveraging compatible data alignment and efficient read/write strategy, \textit{Inhale} can eliminate the majority of unnecessary shift operations with minimal hardware modification. Our evaluation results demonstrate significant throughput-per-area-per-energy (up to 172$\times$) over the state-of-the-art in-memory and ASIC solutions. Future work will extend \textit{Inhale} to provide an end-to-end privacy-preserving solution for IoT devices.

\bibliographystyle{ACM-Reference-Format}
\bibliography{sample-base,google}










\end{document}